\documentclass[amsmath,amssymb,superscriptaddress,prb,nofootinbib,twocolumn]{revtex4-2}
\usepackage{graphicx}
\usepackage{amsmath}
\usepackage{amssymb}
\usepackage{amsfonts}
\usepackage{color}
\usepackage{dsfont}
\usepackage{tikz}
\usepackage{ragged2e}
\usetikzlibrary{calc,decorations.markings}
\usepackage[colorlinks=true,linkcolor=blue,citecolor=blue,urlcolor=blue]{hyperref}
\usepackage{physics}
\usepackage{soul}
\usepackage{float}
\usepackage{subfig}

\newcommand{\x}{\mathsf{x}}

\usetikzlibrary{decorations.pathmorphing, patterns,shapes}
\newcommand{\bla}{bla\\bla\\bla\\bla\\bla}
\definecolor{lime}{HTML}{A6CE39}
\DeclareRobustCommand{\orcidicon}{%
	\begin{tikzpicture}
	\draw[lime, fill=lime] (0,0) 
	circle [radius=0.16] 
	node[white] {{\fontfamily{qag}\selectfont \tiny ID}};
	\draw[white, fill=white] (-0.0625,0.095) 
	circle [radius=0.007];
	\end{tikzpicture}
	\hspace{-2mm}
}
\foreach \x in {A, ..., Z}{%
	\expandafter\xdef\csname orcid\x\endcsname{\noexpand\href{https://orcid.org/\csname orcidauthor\x\endcsname}{\noexpand\orcidicon}}
}
\begin{document}
\title{
Surpassing Carnot efficiency with relativistic motion}
\author{Dimitris Moustos\orcidA{}}
\email{dimitris.moustos@newcastle.ac.uk}
\affiliation{School of Mathematics, Statistics, and Physics, Newcastle University, Newcastle upon Tyne NE1 7RU, United Kingdom}
\author{Obinna Abah\orcidB{}}
\email{obinna.abah@newcastle.ac.uk}
\affiliation{School of Mathematics, Statistics, and Physics, Newcastle University, Newcastle upon Tyne NE1 7RU, United Kingdom}
\date{\today}

\begin{abstract}
Relativistic thermal devices offer a unique platform for understanding the interplay between motion, quantum fields, and thermodynamics, revealing phenomena inaccessible to stationary systems.
We consider a two-qubit SWAP heat engine whose working medium consists of inertially moving Unruh-DeWitt qubit detectors, each coupled to a scalar quantum field in thermal equilibrium at a distinct temperature. Relativistic motion causes the qubits to perceive frequency-dependent effective temperatures that are either hotter or colder than their respective reservoir temperature. We show that the relativistic temperature shift, perhaps the qubit velocity, can be harnessed as a thermodynamic resource to enhance the work output and the efficiency at maximum power of the heat engine. We derive a generalized second law for a heat engine with a moving working medium and demonstrate that it can exceed the standard Carnot bound defined by rest-frame temperatures. 
\end{abstract}
\maketitle

%===========================================================================================================================
%\emph{Introduction.} 
\section{Introduction}
Heat engines are fundamental to the advancement of modern society, from the first industrial revolution to driving today's technologies by converting heat into useful work. The performance of any classical thermal engine is universally constrained by the Carnot limit/efficiency \cite{Callen}, $\eta_C=1-T_c/T_h$, determined solely by the temperatures of the thermal reservoirs, where $T_h$ and $T_c$ denote the  temperatures of the hot and cold reservoirs ($T_h>T_c$), respectively.  Although this bound holds under reversible equilibrium conditions, real machines operate far from such ideals, due to irreversibility and fluctuations. A more practical measure for analyzing the performance of a heat engine is the efficiency at maximum power, popularly called the Curzon-Ahlborn efficiency, $\eta_{CA}=1-\sqrt{T_c/T_h}$ \cite{Curzon:Ahlborn,Deffner:EMP}. However, following the advancement in nanofabrication and coherent control of atoms, the study of thermal machines in the quantum regime has emerged as powerful platforms for investigating the thermodynamic behavior of quantum systems and the fundamental role of information in quantum thermodynamics; see Refs. \cite{vinjanampathy2016quantum,goold2016role,binder2019thermodynamics,Deffner:Campell,Bhattacharjee_2021,therm:eng:rev,potts2024quantumthermodynamics}. The quantum heat engine has been realized in nuclear magnetic resonance \cite{Peterson2019}, an ensemble of nitrogen vacancy center \cite{Klatzow2019} single-ion~\cite{VanHorne2020}, and large quasi-spins \cite{Bouton2021}, with a focus mainly on non-relativistic setups.

Relativistic effects can influence non-equilibrium thermodynamics, from spin–orbit–driven transport in Dirac materials \cite{Werner2019} to light-cone–constrained heat \cite{Pal2020,Paraguass2021} and work fluctuations \cite{Deffner2015}.
Recently, there has been growing interest in understanding the relativistic effects on the performance of quantum thermal devices \cite{Munoz2012,Pena2016,arias2018unruh,myers2021quantum}. 
In particular, attention has been given to thermal machines whose working medium is Unruh-DeWitt (UDW) detectors \cite{Unruh,DeWitt,Birrell:Davies}--qubits interacting with quantum fields while traversing arbitrary trajectories in a background spacetime. An example is the Unruh Otto heat engine \cite{arias2018unruh,gray2018,XU2020135201,Unruh:entangl,barman2022,Mukherjee_2022}, where the detector undergoes uniform acceleration during isochoric strokes and interacts with a quantum field in the vacuum state, effectively acting as a thermal reservoir via the Unruh effect \cite{Unruh}. This framework has been extended to more general accelerated trajectories, such as circular motion, alternative working media such as qutrit detectors, and instantaneous detector-field interactions (see, e.g. \cite{npapadatos,gallock2023quantum,NK:DM,gallock2024relativistic,HT25}). Moreover, studies have examined the impact of relativistic energies \cite{myers2021quantum,chattopadhyay2019relativistic}, as well as the effect of curved spacetime backgrounds--such as black hole spacetimes--on the performance of quantum thermal machines \cite{Ferketic:Deffner,misra2024black,kollas2024,DM:OA} and quantum batteries \cite{DEB,tian2025dissipative,liu2025open}. This raises the fundamental question of whether spacetime geometry can determine the ultimate performance bounds of thermodynamic devices.  

Over the past decades, studies have shown that coupling a heat engine to an engineered nonequilibrium reservoir~\cite{abah2014efficiency}, whether quantum coherent~\cite{Scully2003,Hammam2021NJP}, quantum correlated~\cite{Dillenschneider2009}, or squeezed reservoir~\cite{OAsqueez,Huang:Wang,Manzano2016} can enhance performance and even surpass the standard Carnot bound without violating the second law. The squeezed heat engine has been implemented using a nanomechanical system with reservoirs engineered by
driven squeezed electronic noise \cite{Klaers2017}.

In this manuscript, we present a relativistic quantum heat engine that can outperform classical efficiency bounds without violating the second law. Specifically, we study a quantum SWAP heat engine (two-stroke engine) \cite{Nori2007,campisi2015, TUR:SWAP}, where the two-qubit working medium is modeled by UDW detectors, each locally coupled to a thermal bath at a different temperature. When the qubits move at constant relativistic speed, they can experience frequency-dependent effective temperatures -- potentially hotter or colder than the actual bath temperatures -- which can be exploited to enhance the performance of a quantum heat engine within the universality of the general framework of thermodynamics. Moreover, the setup achieves enhanced efficiency by controlling the motion of the working medium, while both the hot and cold reservoirs remain purely thermal. We remark that the proof-of-principle implementation of a SWAP thermal machine on a quantum processor has demonstrated the use of quantum correlations as a resource to enhance engine performance \cite{Herrera2023}.

%===========================================================================================================================
\section{UDW detector moving through a thermal bath}
%\emph{UDW detector moving through a thermal bath.}
We consider an Unruh-DeWitt detector \cite{Unruh,DeWitt,Birrell:Davies}, modeled as a qubit with energy gap frequency $\omega$ and free Hamiltonian $ H_D\!=\!\hbar\omega\sigma_z/2$, where $\sigma_z$ is the Pauli-Z matrix. The detector is linearly coupled to a massless scalar quantum field $\phi(\x)$ prepared in a thermal state at an inverse temperature $\beta=1/(k_\mathrm{B}T)$, while moving with constant velocity $\upsilon$. The term $k_\mathrm{B}$ is the Boltzmann constant, and hereafter we set the Planck constant $\hbar$ and the speed of light $c$, as $\hbar\!=\!c\!=\!k_B\!=\!1$. The detector worldine is $\x(\tau)=(\gamma\tau,\gamma\upsilon\tau,0,0)$, where $\upsilon$ is the velocity, $\gamma=1/\sqrt{1-\upsilon^2}$ is the Lorentz factor and $\tau$ is the detector's proper time.

In the long-time interaction limit, the qubit is shown to reach a steady state \cite{BJA:DM,Effective:Unruh} described by the reduced density matrix $ \rho_D=e^{-\beta^{\text{eff}}(\omega)H_D}/\text{tr}(e^{-\beta^{\text{eff}}(\omega)H_D})$, $\beta^{\text{eff}}(\omega)$ is the inverse effective temperature $\beta^{\text{eff}}(\omega)=1/T^{\text{eff}}(\omega)$. The frequency-dependent effective temperature reads \cite{matsas95,COSTA1995,Papadatos,verch2025} (see Appendix \ref{appendix} for details):
\begin{align}\label{temp:moving}
T^{\text{eff}}(\omega)= \omega\, \left\{\ln\left(\frac{\ln\left(\frac{1-e^{\beta\gamma(1+\upsilon)\omega}}{1-e^{\beta\gamma(1-\upsilon)\omega}}\right)}{\ln\left(\frac{1-e^{-\beta\gamma(1+\upsilon)\omega}}{1-e^{-\beta\gamma(1-\upsilon)\omega}}\right)}\right)\right\}^{-1},
\end{align}
where $\gamma(1+\upsilon)$ is the blue-shifted and $\gamma(1-\upsilon)$ is the red-shifted Doppler factor.

\begin{figure}[t!]
    \centering   \includegraphics[width=\linewidth]{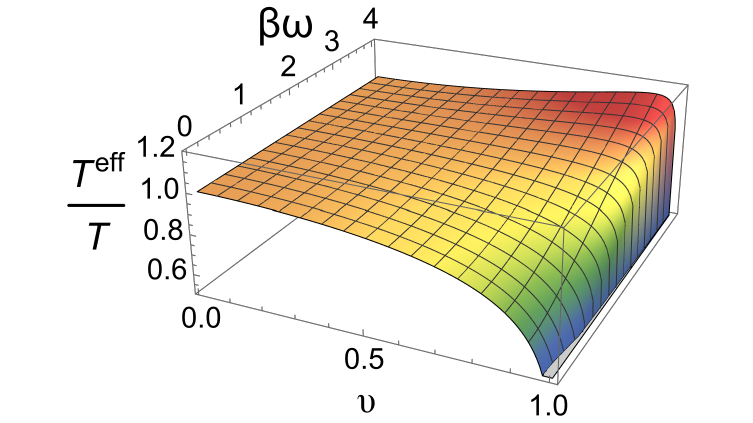}
    \caption{\justifying Effective temperature $T^{\text{eff}}(\omega)$ of a moving qubit scaled to the true temperature $T$ of the heat bath, as a function of the detector's frequency and speed $\upsilon$.}
    \label{fig:effective:temp}
\end{figure}
Figure \ref{fig:effective:temp} shows the effective temperature experienced by the moving qubit detector, normalized to the ambient temperature $T$ of the thermal bath as a function of its frequency and speed. At rest ($\upsilon\to 0$), that is, in the rest frame the qubit temperature is the temperature of the thermal field bath, i.e., $T^{\text{eff}}(\omega)\!=\!T$. However, a moving qubit may perceive either a hotter ($T^{\text{eff}}>T$) or a colder ($T^{\text{eff}}<T$) temperature, depending on both its frequency and the magnitude of its velocity. Specifically, the three possible scenarios are: (i) for small detector frequencies or high-temperature baths ($\beta\omega\ll1$), the effective temperature is always colder than the ambient temperature.  This can be made explicit by expanding the effective temperature for small frequencies, to leading order, giving
\begin{align}\label{teff:series}
    T^{\text{eff}}=\frac{T}{2\gamma\upsilon}\ln\left(\frac{1+\upsilon}{1-\upsilon}\right)+\mathcal{O}(\omega^2).
\end{align}
(ii) On the other hand, for high frequencies or low ambient temperatures ($\beta\omega\gg1$), a moving qubit records an effective temperature that exceeds the rest-frame temperature. 
(iii) In the ultra-relativistic limit ($\upsilon\to 1$), the qubit perceives the thermal field as if it were in its vacuum state, at zero temperature. For a more relevant discussion, see Refs. \cite{matsas95,COSTA1995,Papadatos,verch2025}.

%===========================================================================================================================
%===============================================================================================
\begin{figure}[t!]
    \centering
    \includegraphics[width=\linewidth]{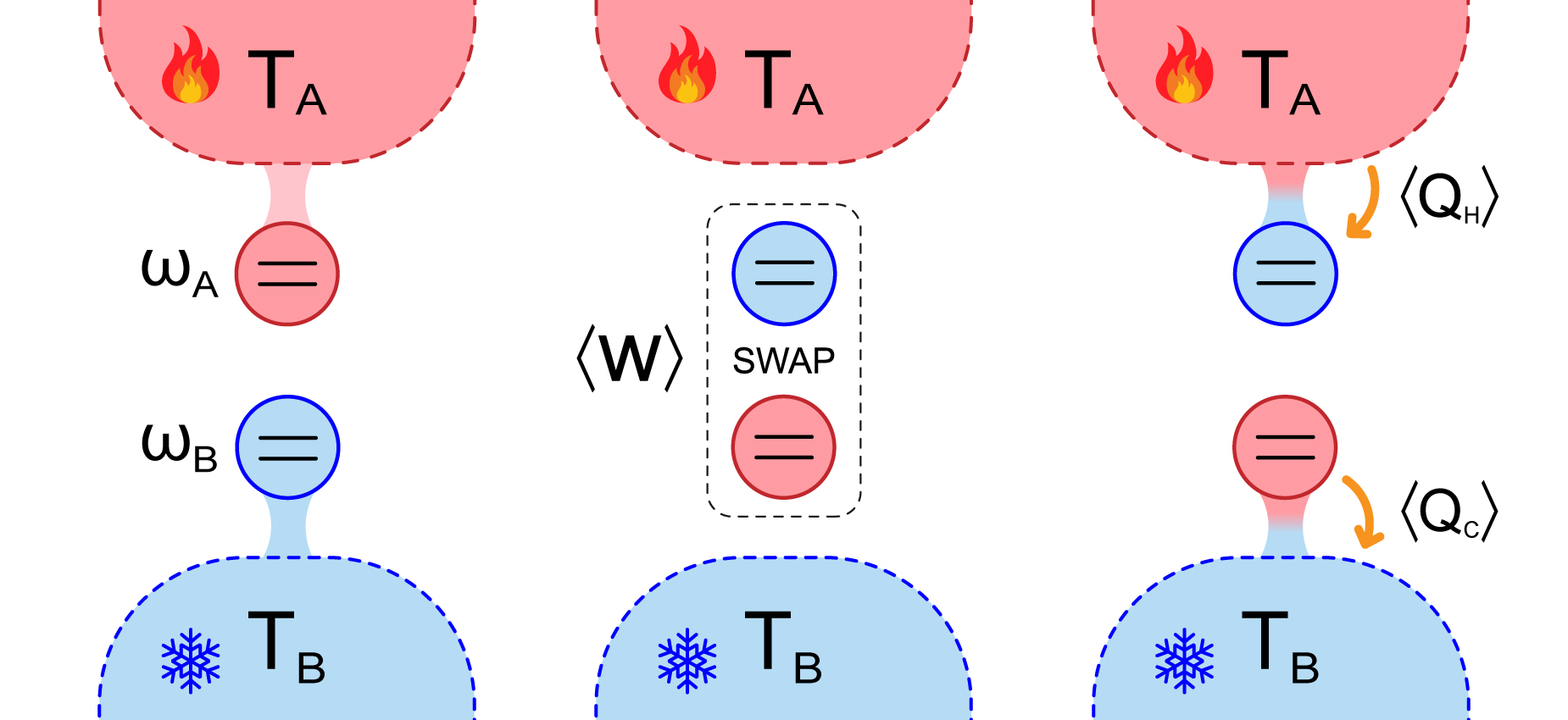}
    \caption{\justifying Schematic representation of a quantum SWAP heat engine. Two qubits are initially brought into thermal equilibrium with separate heat baths at different temperatures $T_A>T_B$. Then, the qubits are decoupled from their respective baths and allowed to interact with each other via a SWAP operation, during which an amount of work $\expval{W}$ is produced.   This procedure can be repeated sequentially,  realizing a two-stroke  heat engine. }
    \label{fig:SWAP}
\end{figure}
%===============================================================================================
\section{Quantum SWAP engine with moving qubits}
%\emph{Quantum SWAP engine with moving qubits.}
We consider a quantum SWAP thermal engine for two qubits that consists of two steps \cite{campisi2015,TUR:SWAP,Herrera2023}, as shown in Figure \ref{fig:SWAP}. The two qubits, $A$ and $B$, with transition frequencies $\omega_A$ and $\omega_B$, are each described by the free Hamiltonian $H_i\!=\!\omega_i\sigma_z^i/2$,
where $i\in\{A,B\}$. 
Initially, qubit $A$ is thermalized with a hot reservoir at temperature $T_A$ and qubit $B$ with a cold reservoir at $T_B$ ($T_A>T_B$). The initial equilibrium states of the composite system is given by $\rho_0=e^{-\beta_AH_A}/Z_A\, \otimes\, e^{-\beta_BH_B}/Z_B$,
with $\beta_i=T_i^{-1}$ and partition function $Z_i\!=\!\text{tr}(-e^{\beta_iH_i})$.

After thermalization, the two qubits are decoupled from their respective thermal baths and allowed to interact with each other via a SWAP unitary, $U\!=\!\frac{1}{2}(1+\boldsymbol{\sigma}_A\cdot\boldsymbol{\sigma}_B)$,  which exchanges their states. Then, the qubits are re-coupled to their respective reservoirs to re-thermalize. Thus, completing a two-stroke quantum engine cycle -- a streamlined analogue of the four-stroke Otto cycle; see \cite{myers2021quantum,binder2019thermodynamics}.

In the case of moving qubits, the two qubits, $A$ and $B$, involved in the implementation of the quantum SWAP engine, are moving with constant speeds $\upsilon_A$ and $\upsilon_B$, respectively. Then, during the first stroke of the engine cycle, the moving qubits $A$ and $B$ interact, with a hot and a cold thermal field bath of temperatures $T_A$ and $T_B$ ($T_A>T_B$) respectively. The relativistic motion modifies the effective temperatures $T^{\text{eff}}_A$ and $T^{\text{eff}}_B$, as defined in Eq. \eqref{temp:moving}. Once each qubit reaches its steady state $\rho_D$, with $D\in\{A,B\}$, the engine proceeds as in the static cycle. We note that, as in the standard UDW
detector framework (e.g., for uniformly accelerated detectors), the relativistic motion of the detectors is assumed to
be externally sustained and their trajectories fixed, without
explicitly modeling the associated energetic cost or back-reaction.

The performance of the heat engine can be characterized in terms of the stochastic variables $Q_H$ and $W$, using the cumulant generating function  $C(\chi_W,\chi_H)$, where $\chi_W$ and $\chi_H$ are the counting fields for work, $W$, and heat exchange with the hot reservoir, $Q_H$, respectively \cite{Schaller2014,Strasberg2022}. 
The cumulants of $Q_H$ and $W$ are obtained by differentiating the cumulant generating function with respect to the corresponding counting fields, as follows;
\begin{align}
   \langle\langle W^m Q_H^{n}\rangle\rangle=(-i)^{m+n}\frac{\partial^{m+n}C(\chi_W,\chi_H)}{\partial\chi^m_W\partial\chi^n_H}\Bigg|_{\chi_W=\chi_H=0}.
\end{align}
Using the two-point measurement scheme \cite{RevTPM} to jointly estimate $Q_H$ and $W$, the characteristic function takes the form \cite{TUR:SWAP,Sacchi2021A,Sacchi2021}:
\begin{align}
    C(\chi_H,\chi_W)\!=&\!\ln\bigg\{\!\text{tr}\bigg[ U^\dagger\!\left(e^{i(\chi_W-\chi_H)H_A} e^{i\chi_WH_B}\right)U\nonumber\\&\!\left(\!e^{-i(\chi_W-\chi_H)H_A} e^{-i\chi_WH_B}\!\right)\rho_A\otimes\rho_B\bigg]\!\bigg\},
\end{align}
from which we obtain the mean work and input heat,
\begin{align}\label{work:average}
    \expval{W}\!=\!\frac{\!\omega_B\!-\!\omega_A\!}{2}\bigg(\!\tanh(\frac{\beta^{\text{eff}}_B\omega_B}{2})\!-\!\tanh(\frac{\beta^{\text{eff}}_A\omega_A}{2})\!\bigg),
\end{align}
\begin{align}
    \expval{Q_H}
    \!=\!\frac{\omega_A}{2}\bigg(\!\tanh(\frac{\beta^{\text{eff}}_B\omega_B}{2})\!-\!\tanh(\frac{\beta^{\text{eff}}_A\omega_A}{2})\!\bigg).
\end{align}
The relativistic two-stroke quantum SWAP thermal cycle functions as a heat engine when the total work done is positive. The modified positive work condition is $\beta^{\text{eff}}_A/\beta^{\text{eff}}_B<\omega_B/\omega_A<1$.

%===========================================================================================================================
\begin{figure}[tp]
    \centering    \includegraphics[width=0.485\linewidth]{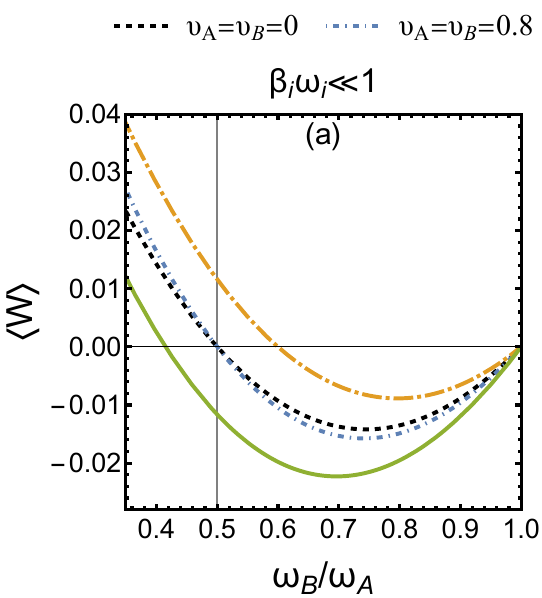}\hspace{0.16cm}\includegraphics[width=0.495\linewidth]{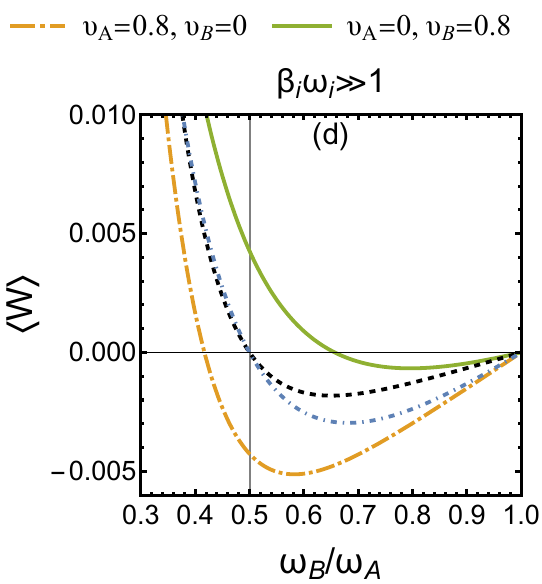}\\   \vspace{0.18cm}\includegraphics[width=0.485\linewidth]{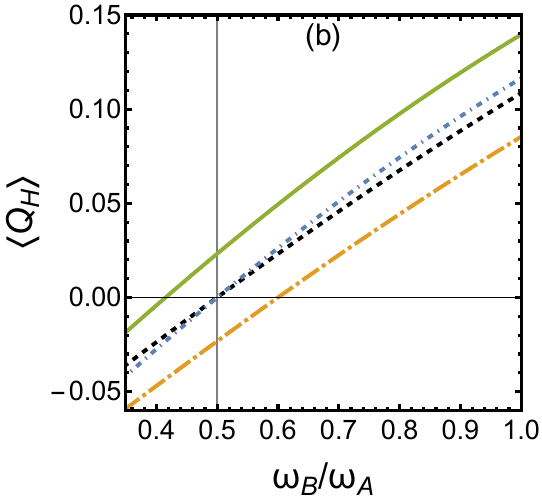}\hspace{0.16cm}\includegraphics[width=0.485\linewidth]{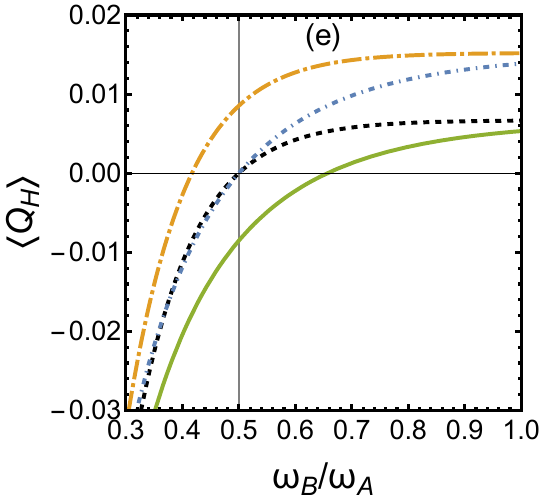}\\ \vspace{0.18cm}   \includegraphics[width=0.485\linewidth]{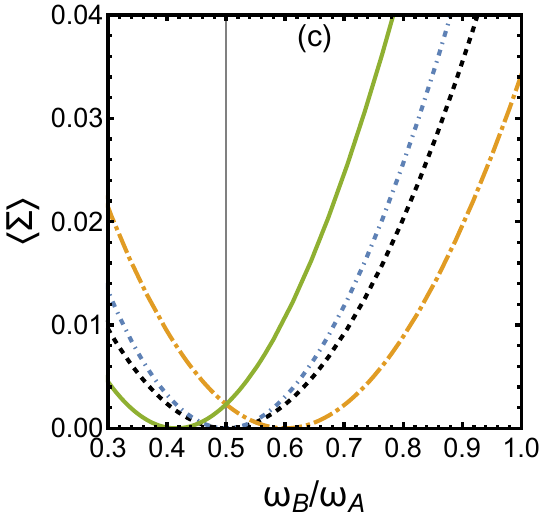}\hspace{0.16cm}\includegraphics[width=0.485\linewidth]{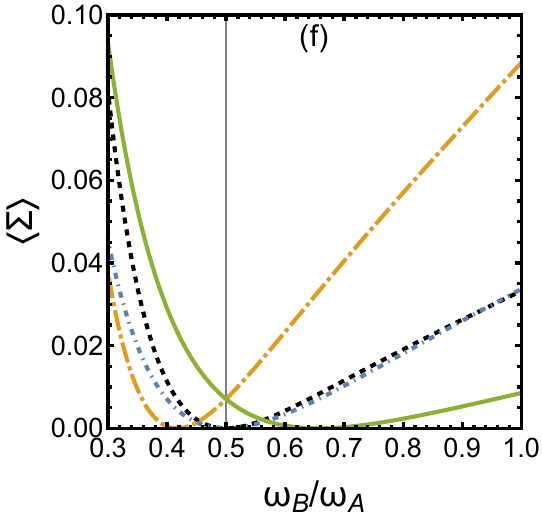}
    \caption{\justifying  Averages of work $\expval{W}$, heat exchanged with the hot bath $\expval{Q_H}$, and entropy production $\expval{\Sigma}$ are shown as functions of the frequency ratio $\omega_B/\omega_A$, for a fixed temperature ratio of the thermal baths $\beta_A/\beta_B = 1/2$ and varying qubit speeds. The left panels corresponds to the high-temperature/low-frequency regime ($\beta_i\omega_i\ll1$), while the right panels to the low-temperature/high-frequency regime ($\beta_i\omega_i\gg1$). The vertical line indicates the boundary between refrigerator and heat engine operational regimes in the rest-frame case.}
    \label{fig:variables}
\end{figure}
%===========================================================================================================================

In Figure \ref{fig:variables}, we present the average values of work $\expval{W}$, and heat exchanged with the hot bath $\expval{Q_H}$, as functions of the frequency ratio $\omega_B/\omega_A$, for a fixed temperature ratio of the thermal baths $\beta_A/\beta_B\!=\!1/2$. In the regimes of high-temperature/low-frequency (Fig. \ref{fig:variables} (a)-(c)) and low-temperature/high-frequency (Fig. \ref{fig:variables} (d)-(f)), we examine three different scenarios: (i) both qubits $A$ and $B$ move through their respective thermal reservoirs with the same relativistic velocity $\upsilon_A\!=\!\upsilon_B$ (ii) only qubit $A$ moves through the hot reservoir; and (iii) only qubit $B$ moves through the cold bath. These are compared with the case where both qubits remain at rest, that is, a standard quantum SWAP heat engine.

In the high temperature regime, the leading order expansion of the effective temperature Eq. \eqref{teff:series} shows that when qubit $B$ moves with speed $\upsilon_B=0.8$ while qubit $A$ remains at rest,  the device operates as a heat engine when $0.42\lesssim \omega_B/\omega_A<1$. This regime is consistent with the values illustrated in Fig. \ref{fig:variables}. It indicates that the device can function as a heat engine in parameter regimes where, in the static case, it would act as a refrigerator.

Moreover, in the high-temperature regime, the total work output $\expval{W}$ of the heat engine increases when the qubit $B$ moves at relativistic speed through the cold reservoir, while the qubit $A$ remains static, compared to the standard case where both qubits are at rest. In contrast, in the low-temperature regime, the work output increases when qubit A moves through the hot bath while qubit B remains at rest. This behavior can be understood by means of Eq. \eqref{work:average}: the qubit moving through the cold (hot) reservoir perceives an effective temperature lower (higher) than the temperature of the rest frame, which leads to an enhanced work output.

The efficiency of the heat engine $\eta$, defined as the ratio of the total work output to the absorbed heat, reads
\begin{align}
     \eta=-\frac{ \expval{W}}{ \expval{Q_H}}=1-\frac{\omega_B}{\omega_A}.
     \label{efficiency}
\end{align}
Based on the second law of thermodynamics, the entropy production $\expval{\Sigma}$ of the heat engine is always positive (see, e.g., Fig. \ref{fig:variables} and Appendix \ref{app:B}) and evaluated as \cite{abah2014efficiency,Landi2021}, 
\begin{align}
    \expval{\Sigma}=(\beta^{\text{eff}}_B-\beta^{\text{eff}}_A)\expval{Q_H}+\beta^{\text{eff}}_B\expval{W}\ge 0.
\end{align}
Therefore, the maximum efficiency of the heat engine is bounded by the generalized Carnot efficiency, $\eta_C^{\text{eff}}$, \cite{abah2014efficiency}
\begin{align}\label{Gen:CarnEffic}
\eta \leq 1-\frac{T^{\text{eff}}_B}{T^{\text{eff}}_A} := \eta_C^{\text{eff}}.   
\end{align}

%=================================================================================
\section{Efficiency at maximum power.}
%\emph{Efficiency at maximum power.}  
A useful quantity to characterize the performance of a heat engine is the efficiency at maximum power \cite{andresen2011}. This is determined by optimizing the power output per unit time 
\label{power:out}
   $ \!P\!=-\expval{W}_{\text{tot}}/\tau$,
with respect to a system parameter, and then evaluating the corresponding efficiency at that optimized power using Eq. (\ref{efficiency}). We separately analyze the high-temperature and low-temperature regimes.
%=========================================================================================================================================
\begin{figure}[tp]
    \centering 
     \centering    \includegraphics[width=0.485\linewidth]{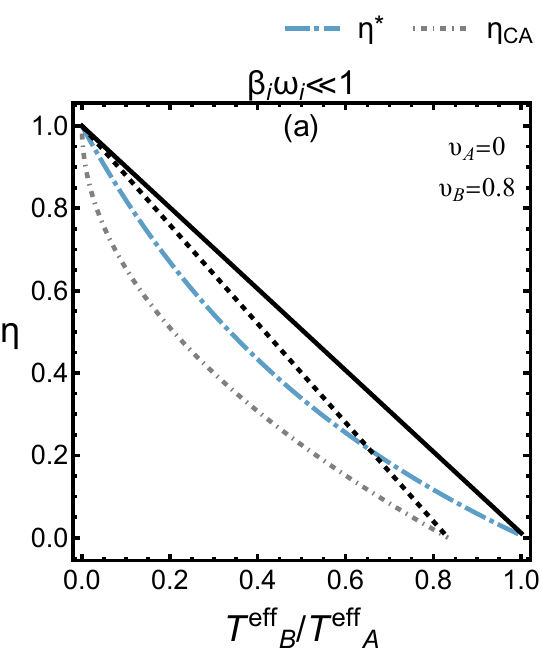}\hspace{0.16cm}\includegraphics[width=0.485\linewidth]{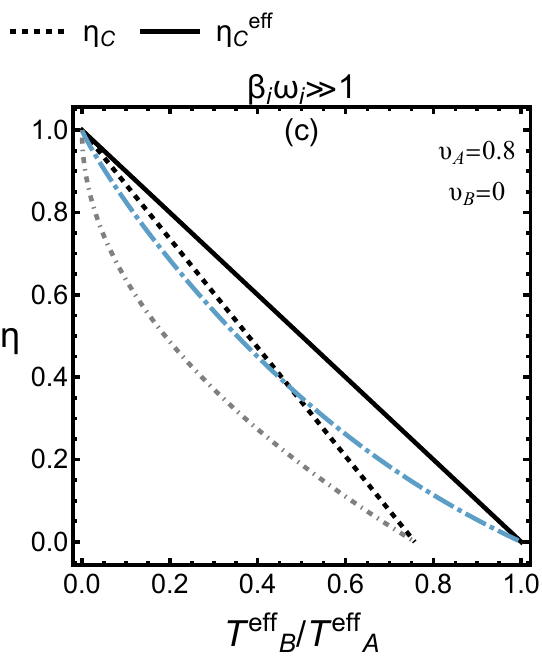}\\   \vspace{0.18cm}\includegraphics[width=0.485\linewidth]{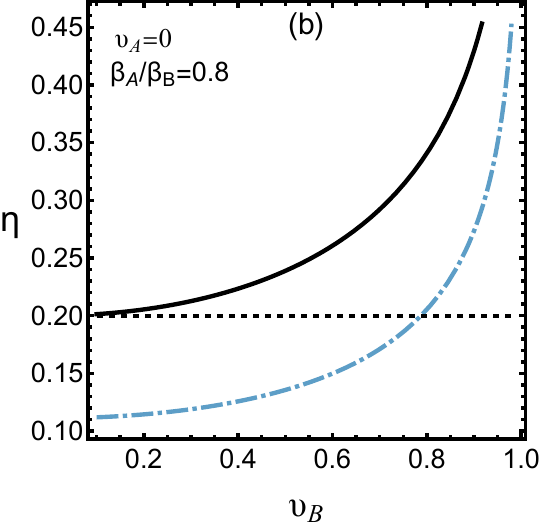}\hspace{0.16cm}\includegraphics[width=0.475\linewidth]{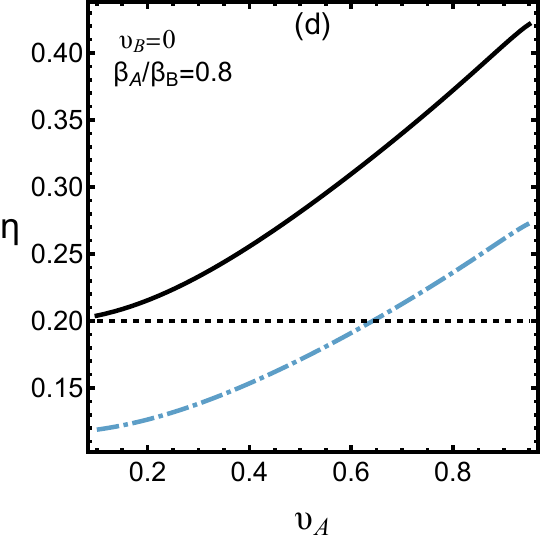}
    \caption{\justifying Efficiency of the quantum heat engine as function of the effective temperature ratio $T^{\text{eff}}_B/T^{\text{eff}}_A$ (for fixed qubit velocities), and  as a functions of the velocity of qubit $B$ and qubit $A$ (for fixed ambient temperature ratios). The left panels represent the high-temperature/low-frequency regime ($\beta_B\omega_B=0.12$), while the right panels show the low-temperature/high-frequency regime ($\beta_A\omega_A\!=\!6.5$).
     Here, $\eta^*$ denotes the efficiency at maximum power;  $\eta_{CA}$ is the Curzon–Ahlborn efficiency; $\eta_C$ is the standard Carnot efficiency; and $\eta_C^{\text{eff}}$ is the generalized Carnot bound.}
    \label{fig:performance}
\end{figure}
%=========================================================================================================================================
In the high-temperature regime, the power output of the relativistic heat engine is optimized with respect to frequency $\omega_A$, while keeping the temperatures $\beta_A$, $\beta_B$, and frequency $\omega_B$ fixed. We find that the power output is maximized when the qubit frequencies satisfy the condition $\omega_B/\omega_A=2\beta^{\text{eff}}_A/(\beta^{\text{eff}}_A+\beta_B^{\text{eff}})$. The resulting efficiency at maximum power is
\begin{align}
    \eta^*=1-2\frac{\beta^{\text{eff}}_A}{\beta^{\text{eff}}_A+\beta^{\text{eff}}_B}%=\frac{\eta_C^{\text{eff}}}{2-\eta_C^{\text{eff}}}
    ,\label{EMP_high}
\end{align}
where $\beta^{\text{eff}}_{i}$ is the relativistic motion dependent temperature defined in Eq. \eqref{teff:series}. In constant velocities, Eq.~\eqref{EMP_high} reduces to the efficiency at the maximum power of the two-level quantum Otto engine \cite{singh:OA}.  In the low-temperature regime, by optimizing the power output with respect to $\omega_B$, we find that power is maximized when the qubit frequencies satisfy the condition
\begin{align}
    \omega_B=\omega_A+\left(1-\ W\left(e^{\omega_A(\beta^{\text{eff}}_B-\beta^{\text{eff}}_A)+1}\right)\right)\Big/\beta_A^{\text{eff}},
\end{align}
where $W(x)$ stands for the Lambert function \cite{NIST}. Figure~\ref{fig:performance} shows how the efficiency at maximum power of the heat engine varies across high- and low-temperature regimes -- increasing or decreasing with the effective temperature ratio $T^{\text{eff}}_B/T^{\text{eff}}_A$ (for fixed qubit velocities), and showing the dependence on the velocity of qubit $A$ and qubit $B$ (for fixed true temperature ratios).Figure~\ref{fig:performance} (a) and (b) suggest that a finite $\eta$ can be observed even when the bath temperatures are equal, $T_A=T_B$, in which case the standard Carnot bound vanishes. This demonstrates that relativistic motion itself can serve as a thermodynamic resource.

In the high temperature, we find that, when qubit $B$ is moving through the cold reservoir with constant relativistic speed -- thus perceiving an effective temperature $T_{B}^{\text{eff}}<T_B$-- it achieves efficiency at maximum power $\eta^*$ that exceeds the Curzon–Ahlborn efficiency $\eta_{CA}$, and can even surpass the Carnot efficiency $\eta_{C}$ defined with respect to the rest-frame temperatures $T_A$ and $T_B$. Analogous behavior occurs in the low-temperature regime when qubit $A$ moves through the hot reservoir, such that experiencing an effective temperature $T_{A}^{\text{eff}}>T_A$. We note that Fig. \ref{fig:variables} suggests that the work output can potentially increase when both qubits move at the same velocity. However, it can be directly seen by combining Eqs. \eqref{temp:moving} and \eqref{Gen:CarnEffic} that the engine's efficiency in this case always remains bounded by the standard Carnot limit, $\eta_C$. These observations indicate that positive work can be extracted even when $T_B>T_A$, provided that $T_{B}^{\text{eff}}/T_{A}^{\text{eff}}<1$ -- a behavior reminiscent of the enhancement of thermal engine performance in quantum Otto heat engines with squeezed thermal reservoirs \cite{abah2014efficiency,Huang:Wang}.

%========================================
%========================================

\section{Conclusions}
%\emph{Conclusions.} 
We have shown that the performance of a quantum thermal engine can be enhanced by the relativistic motion of the engine working medium. The efficiency at maximum power may exceed the Carnot limit defined by the rest-frame bath temperatures without violating the second law of thermodynamics. This improved performance stems from the fact that the moving working medium (qubits) perceives shifted effective temperatures -- hotter or colder than the baths in their rest frames -- reflecting the lack of a unique Lorentz transformation rule for temperature \cite{matsas95,COSTA1995,landsberg1996laying,Papadatos,verch2025}.
Our findings highlight that harnessing the relativistic effects opens new pathways to manipulate energy flow and design more energy efficient quantum thermal devices.

%========================================
%========================================
\section{Acknowledgments}
%\emph{Acknowledgments.}
D. M. wishes to thank Charis Anastopoulos for fruitful discussions. This work was supported by the UK Research and Innovation Engineering and Physical Sciences Research Council (Grant No. EP/Z002796/1).
%===========================================================================================

\appendix
\section{Effective temperatures recorded by UDW detectors}\label{appendix}
We consider an Unruh-DeWitt detector \cite{Unruh,DeWitt,Birrell:Davies}, modeled as a qubit with a frequency $\omega$ between its two energy levels, and free Hamiltonian $ H_D=\omega\sigma_z/2$.
The detector moves along a trajectory $\mathsf{x}(\tau)$ parametrized by its proper time $\tau$, and is linearly coupled to a massless scalar quantum field $\phi(\x)$. In the interaction picture, the Hamiltonian that describes the interaction between the detector and the field reads
\begin{align}\label{UDW:Hamilt}
H_{\text{int}}(\tau)=\lambda \mu(\tau)\phi(\mathsf{x}(\tau)),
\end{align}
where $\lambda$ is a coupling constant, $\mu(\tau)=e^{i\omega\tau}\sigma_++e^{-i\omega\tau}\sigma_-$ is the detector's monopole moment operator expressed in
terms of the Pauli ladder operators $\sigma_+$ and $\sigma_-$, and $\phi(\mathsf{x}(\tau))$ is the field evaluated along the detector’s trajectory.

We now consider a detector moving along a stationary trajectory \cite{Letaw}--a special case of which is the inertial
motion--interacting with a quantum field prepared in a thermal state at temperature $T=\beta^{-1}$. In this case, the pullback of the field's Wightman two-point correlation function along the detector's trajectory, defined as $\mathcal{W}(\tau,\tau'):=\expval{\phi(\x(\tau))\phi(\x(\tau'))}_{\beta}$, is stationary; that is, it depends only on the proper time deference $\tau-\tau'$ between two points on the detectors' worldline, and can be expressed as $\mathcal{W}(\tau,\tau')=\mathcal{W}(\tau-\tau')$. The detector’s transition rate \cite{Birrell:Davies}—the probability per unit proper time for the detector to transition between its energy levels due to interaction with the field—is then given by the Fourier transform of the Wightman function
\begin{align}\label{transition:rate}
    \mathcal{G} (\omega):=\int_{-\infty}^{+\infty}ds\,e^{-i\omega s}\mathcal{W}(s).
\end{align}

By means of the generalized detailed balance condition \cite{fewster2016}, $\mathcal{G}(-\omega)=e^{\omega/T^{\text{eff}}(\omega)}\mathcal{G}(\omega)$, which relates the excitation and de-excitation rates of the detector, one can define a frequency-dependent \emph{effective temperature} as
\begin{align}
    T^{\text{eff}}(\omega)=\omega\bigg/\ln\left(\frac{\mathcal{G}(-\omega)}{\mathcal{G}(\omega)}\right),
\end{align}
interpreted as the temperature experienced by the moving detector \cite{Effective:Unruh}. In the long-time interaction limit, the detector relaxes to a steady state described by the reduced density matrix \cite{BJA:DM}:
\begin{align}\label{app:dm}
    \rho_D=\frac{e^{-\beta^{\text{eff}}(\omega)H_D}}{\text{tr}(e^{-\beta^{\text{eff}(\omega)}H_D})},
\end{align}
where $\beta^{\text{eff}}(\omega)=1/T^{\text{eff}}(\omega)$ in the inverse effective temperature.

Now, we consider the case of a detector moving with constant velocity $\upsilon$, through the thermal field bath. The detector's worldline is $\x(\tau)=(\gamma\tau,\gamma\upsilon\tau,0,0)$, where $\gamma=1/\sqrt{1-\upsilon^2}$ is the Lorentz factor. In this case, the detector's transition rate takes the form \cite{matsas95,COSTA1995,Papadatos} (see also \cite{verch2025} for a detailed derivation):
\begin{align}
     \mathcal{G} (\omega)=\frac{\lambda^2}{4\pi\beta\gamma\upsilon}\ln\left(\frac{1-e^{-\beta\gamma(1+\upsilon)\omega}}{1-e^{-\beta\gamma(1-\upsilon)\omega}}\right).
\end{align}
In the asymptotic limit, the detector's reduced density matrix is of the form \eqref{app:dm}, with an effective temperature given by 
\begin{align}
T^{\text{eff}}(\omega)=\omega\bigg/\ln\left(\frac{\ln\left(\frac{1-e^{\beta\gamma(1+\upsilon)\omega}}{1-e^{\beta\gamma(1-\upsilon)\omega}}\right)}{\ln\left(\frac{1-e^{-\beta\gamma(1+\upsilon)\omega}}{1-e^{-\beta\gamma(1-\upsilon)\omega}}\right)}\right).
\end{align}

\onecolumngrid
\section{The 2nd law}\label{app:B}
The cumulant generating function,  $C(\chi_W,\chi_H)=\ln\expval{e^{i\chi_WW+i\chi_HQ_H}}$, explicitly reads
\begin{align}
 C(\chi_W,\chi_H)=\ln\left(\frac{\cosh\bigg(\frac{1}{2}(\beta^{\text{eff}}_A\omega_A+i(\omega_A(\chi_W-\chi_H)+\omega_B\chi_W))\bigg)\cosh\bigg(\frac{1}{2}(\beta^{\text{eff}}_B\omega_B-i(\omega_A(\chi_W-\chi_H)+\omega_B\chi_W))\bigg)}{\cosh(\beta^{\text{eff}}_A\omega_A/2)\cosh(\beta^{\text{eff}}_B\omega_B/2)}\right).
\end{align}
 From this form, it is straightforward to verify the identity $ C(i\beta^{\text{eff}}_B,i(\beta^{\text{eff}}_B-\beta^{\text{eff}}_A))=0$, leading to fluctuation theorem $\expval{e^{-\Sigma}}=1$ . When combined with the Jensen's inequality, this implies the second law of thermodynamics 
$\expval{\Sigma}\geq0$.
%=============================================================================================
\twocolumngrid

\bibliography{references}

\end{document}